\documentstyle[12pt]{article}
\if@twoside  m
\else
\fi

\title{Viscosity of Quantum Hall Fluids}
\author{J.~E.~Avron\\ Department of Physics, Technion, 32000 Haifa,
Israel \and R.~Seiler and P.~G.~Zograf \,$^*$\\  Fachbereich Mathematik,
Technische Universit\"at Berlin, 10623 Berlin}

\begin{document}

\maketitle

\begin{abstract}
\noindent
The viscosity of quantum fluids with an energy gap at zero temperature is
non-dissipative and is related to  the adiabatic curvature on the space of
flat background metrics (which plays the role of the parameter space).
For a quantum Hall fluid on  two dimensional tori this viscosity is computed.
In this case the average viscosity is quantized and is proportional to
the total magnetic flux through the torus.
\end{abstract}

\vfill
PACS: 72.10.Bg, 67.50.-b

{\it{ \footnotesize ${}^*$  On leave from Steklov Mathematical Institute,
St. Petersburg, Russia}}

\newpage

Classically, the  elastic modulus $\lambda$ and
viscosity $\eta$ are tensors of rank 4, which relate the stress tensor $T$
to the strain tensor $e$ and strain-rate tensor $\dot e$
\cite{Landau-Lifshitz}.  In the limit of small strain-rates
\begin{eqnarray}T_{\alpha\beta}(e, \dot e) =
T_{\alpha\beta}(e, 0) + \sum_{\gamma,\delta}
\eta_{\alpha\beta\gamma\delta}\ \dot e_{\gamma\delta},
\quad\alpha,\beta,\gamma,\delta =1,\dots,d, \label{c-transport}\end{eqnarray}
where d is the dimension of configuration space. The elastic modulus
tensor \break \hbox{$\lambda_{\alpha\beta\gamma\delta}= -\frac{1}{2}\
\frac{\partial T_{\alpha\beta}(e, 0)}{\partial e_{\gamma\delta}}$} is symmetric
in all its indices.  Classical Newtonian fluids \cite{Landau}
offer no resistance to shear; so,  all shear modes are
soft and $\lambda$ has rank one.

The term $\sum \eta_{\alpha\beta\gamma\delta}\ \dot e_{\gamma\delta}$
in Eq.~(\ref{c-transport}) is {\em
the viscosity stress tensor} \cite{Landau}. For Newtonian fluids, $T$ and
$\dot e$ are symmetric tensors, and consequently,
$\eta_{\alpha\beta\gamma\delta}=\eta_{\beta\alpha\gamma\delta}=
\eta_{\alpha\beta\delta\gamma}$ \cite{Landau-Lifshitz,deGennes}.
With respect to substitution of indices $(\alpha\beta\gamma\delta)\mapsto
(\gamma\delta\alpha\beta)$, the viscosity tensor
splits into symmetric and  anti-symmetric parts, $\eta=\eta^S+\eta^A$, where
\begin{eqnarray} \eta^S_{\alpha\beta\gamma\delta} =
\eta^S_{\gamma\delta\alpha\beta},\quad
\eta^A_{\alpha\beta\gamma\delta}=-\eta^A_{\gamma\delta\alpha\beta}.
\end{eqnarray}  The symmetric part, being associated with dissipation, is
a definite quadratic form on the space of 2-tensors (strain-rates).
If a fluid is isotropic, then
\begin{eqnarray} \eta^S_{\alpha\beta\gamma\delta}\ = \eta \
\delta_{\alpha\gamma} \,\delta_{\beta\delta}+ \xi \delta_{\alpha\beta}
\,\delta_{\gamma\delta}, \end{eqnarray} where $\xi$ and $\eta$ are the two
{\em
viscosity constants} \cite{Landau}. If the fluid is also incompressible,
$\xi$ may be put to
zero with no loss. This is the normal situation in Newtonian fluid mechanics.
The  anti-symmetric part of the viscosity tensor describes non-dissipative
response. One normally assumes that it is absent
because of no compelling evidence to
think otherwise.

 Quantum fluids can have a ground state which is
separated  by a finite gap from the rest of the spectrum.  At zero temperature
such a fluid will have a non-dissipative response, with $\eta^S=0$. The
anti-symmetric part $\eta^A$ may or may not vanish at zero temperature.
For instance, $\eta^A$ vanishes for systems with time reversal symmetry
 (and this may be viewed as a consequence of Onsager relation
\cite{Callen}).

Quantum fluids with energy gap and broken time reversal symmetry will,
in general, have
$\eta^A\neq 0$ at zero temperature. A quantum Hall fluid with a full Landau
level
gives such an example. The viscosity tensor at low temperature could then be
dominated by the non-dissipative part $\eta^A$.   The study of the
non-dissipative viscosity bears analogies with the Quantum Hall effect
\cite{Stone}, the Magnus force in superconductors \cite{Thouless-Ao}, and even
with gravity \cite{FS}.  Like the Hall conductance, and the Magnus force, the
(non-dissipative) viscosity is related to the adiabatic curvature \cite{Berry}
and to topological invariants \cite{TKNN}.  The connection with gravity comes
about because the adiabatic curvature relevant for viscosity is a 2-form on
the space parameterizing flat background metrics. In other words,
the viscosity stress tensor  can be viewed as describing the response of the
system to certain adiabatic deformations of the metric.

Let us first recall a general fact from the theory of adiabatic response
\cite{Adiabatic}. Consider a family of Hamiltonians $H(X)$ which
depend smoothly on a set of parameters $X=\{X_1,\dots,X_n\}$ ( $X$
denotes a point in parameter space while $x$ denotes a point in configuration
space). Let  $|\psi(X)\rangle$ be a (normalized)
non-degenerate state of $H(X)$, with
energy  $E(X)$. Let $X(t)$ be a path in parameter space which is traversed
adiabatically;  $\dot X$ is the velocity along the path. We
assume throughout that the state stays non-degenerate along the path.

By the principle of virtual work,
$-\frac{\delta H}{\delta X}$  is the observable corresponding to the
generalized force related to $\delta X$ \cite{Feynman}.
Adiabatic response theory says that, in the adiabatic limit,
\begin{eqnarray}
\langle \frac{\partial H}{\partial X_j}\rangle =
\frac{\partial E}{\partial X_j} + \sum_{k=1}^n \Omega_{jk} \, \dot
X_k,\label{basic-transport}
\end{eqnarray}
where $\Omega_{ij}$ is the (anti-symmetric) adiabatic curvature \cite{Berry}:
\begin{eqnarray}
\Omega_{ij} \equiv \, Im\  \langle
\partial_i\psi|\partial_j\psi\rangle, \quad
\Omega_{ij}= -\Omega_{ji}, \quad
\partial_j = \frac{\partial}{\partial X_j}.
\end{eqnarray}
If $H$ is time reversal invariant, then $\Omega_{ij}=0$ \cite{Berry}. The
anti-symmetry of $\Omega$ implies no dissipation: there is no change in energy
if the system is taken along a closed loop in parameter space.

In the special case where $|\psi\rangle$ is a (normalized) multiparticle
state
corresponding to N non-interacting fermions in the single particle (normalized)
states  $|\varphi_\ell\rangle\ $  with energies $E_\ell (X)$
(all depending smoothly on parameter $X$) one has
\begin{eqnarray} \Omega_{jk}= \, Im\ \sum_{\ell=1}^N \  \langle \partial_j
\varphi_\ell|\partial_k \varphi_\ell\rangle, \quad E(X)= \sum_{\ell=1}^N\
E_\ell(X).\label{fermions} \end{eqnarray}

To apply all this to viscosity one needs to identify appropriate parameters so
that $ \frac{\partial H}{\partial X_j}$ is related to the
energy-momentum tensor  and $\dot X$ to the rate of strain. We shall argue
that this
is done by flat metrics (metrics  with constant coefficients).

Consider a fluid confined to a given domain in  d dimensional Euclidean space.
The planar parallelogram shown in Fig.~1 is an example for $d=2$. A uniform
deformation is represented by the (symmetric) tensor of constant strain
\cite{Landau-Lifshitz}   \begin{eqnarray} e_{\alpha\beta} =
\frac{1}{2}\left(-\delta_{\alpha\beta}+\sum_\gamma\,  \frac {\partial
x'_\gamma}{\partial x_\alpha} \frac {\partial x'_\gamma}{\partial
x_\beta}\right),  \end{eqnarray}
 where $x'$ is a linear transformation of
$x$. In two dimensions  the deformation space of flat metrics is three
dimensional.  A two dimensional subspace is associated with transformations
that
preserve the volume, and  the transverse direction may be associated with
scaling.

One  can view  deformations  as changing the domain while keeping the
underlying Euclidean metric fixed. An equivalent point of view is to
consider a deformation of the metric,  while keeping the domain fixed with
$g=I+2e$  \cite{Landau-Lifshitz}. The deformed metric is not Euclidean in
general but always has constant coefficients  and therefore flat. Convenient
coordinates on the corresponding parameter space will be introduced later.

Let $H(g)$ denote the Hamiltonian on a domain $D$ in d-dimensional space
associated with metric tensor $g$
with constant coefficients $g_{\alpha\beta}$ viewed as external
parameters.  By the
principle  of virtual work \cite{Feynman},
$-\frac{\delta H}{\delta e} =-2\ \frac{\delta H}{\delta g} $  is
the observable associated with {\em total} stress tensor, $ \int_D
T_{\alpha\beta}(x)\ d\,vol(x)$.  (The total
stress has dimension of energy, while the stress has the dimension of
pressure).
For example,  for a free particle of mass m in $D$ the kinetic energy is
\begin{eqnarray} H(g) = \frac{m}{2}\ \sum\, g^{\alpha\beta}\, v_\alpha v_\beta
,\end{eqnarray} and the total stress is the observable
\begin{eqnarray}-\frac{\partial H}{\partial e_{\alpha\beta}} \Big|_{e=0}=
m\  v_\alpha  v_\beta.\end{eqnarray}
This is reminiscent of  general relativity where the (local) energy-momentum
tensor is related to the variation of the action due to
(local) deformation of the metric.

Adiabatic deformations of the strain give the quantum version of Eq.
(\ref{c-transport}):
 \begin{eqnarray}
\langle \frac{\partial H}{\partial e_{\alpha\beta}}\rangle= \frac{\partial E}
{\partial e_{\alpha\beta}} +
\sum_{\gamma,\delta=1}^d \Omega_{\alpha\beta\gamma\delta} \, \dot
e_{\gamma\delta}.\label{q-transport}
\end{eqnarray}
The elastic modulus tensor $\lambda_{\alpha\beta\gamma\delta}= \frac{1}{2}\
\frac{\partial^2 \, E}{\partial e_{\alpha\beta}\,\partial e_{\gamma\delta}}$
is manifestly symmetric in all its indices. The adiabatic curvature plays
here the role of non-dissipative viscosity. For homogeneous fluids the
viscosity tensor and the adiabatic curvature are related by \begin{eqnarray}
\Omega=V\ \eta^A,\end{eqnarray} where $V$ is the volume of the fluid.
The adiabatic curvature $\Omega$ has
units of Planck constant $\hbar$. This is analogous to the situation in the
quantum Hall effect where the adiabatic curvature (=conductance) has units of
$e^2/\hbar$.

Let us illustrate these ideas on a concrete example.
Consider two dimensional Quantum Hall fluid  on a torus, such as in
Fig.1, with
opposite sides identified. Explicitly, let  $Q$ be the unit square in
configuration space: $(x,y)\in Q =[0,1]\times[0,1]$.
We shall choose the following coordinates on the
deformation space of flat metrics:
$\{V,\tau_1,\tau_2\ |\ V,\tau_2 >0\}$. By $V= \sqrt{\det g}$
we denote the area and we use
complex variable $\tau=\tau_1+i\,\tau_2$ to parameterize
 deformations that preserve the volume.

The flat metrics $g(V,\tau)$ are:
\begin{eqnarray}
g(V,\tau)=\frac{V}{\tau_2} (dx^2 +2 \tau_1 dx\,dy
+|\tau|^2\,dy^2),\quad x,y\in Q.\label{metric}
\end{eqnarray}
The Landau Hamiltonian  describes the kinetic energy of a charged (spinless)
particle in  a constant magnetic field, and Aharonov-Bohm gauge fields. It is
given by
\begin{eqnarray}  H(V,\tau, \phi) = \frac{1}{V\tau_2}\left( |\tau|^2
D_x^2 -  \tau_1\,(D_xD_y+D_yD_x) +D_y^2\right),\label{hamiltonian}
\end{eqnarray}
where
$D_x=-i\partial_x+2\pi( B y+\phi_1+B/2)$, $D_y=-i\partial_y+2\pi(\phi_2+B/2)$;
$\phi_1$ and $\phi_2$ are associated with two Aharonov-Bohm fluxes,
and  the integer $B$ is the number of magnetic flux quanta through torus. (Note
that in our units $1$ is the unit of quantum flux, $hc/e$.)
We impose the usual magnetic translation boundary conditions \cite{Zak}:
\begin{eqnarray} \psi(x+1,y)=\psi(x,y),
\quad \psi(x,y+1)= e^{-2\pi i Bx}\,\psi(x,y). \end{eqnarray}
The (single particle) ground state is $B$-fold
degenerate  with energy  $E =2\pi B/V$ independent of $\tau$ and $\phi$.
The ground state of a full Landau level has energy $E=2\pi B^2/V$ and is
separated by a gap from the rest of the spectrum. (For a non-relativistic
spinning electron, described by Pauli equation, one has $E=0$). It follows from
Eq. (\ref{q-transport}) that the only non-zero component of the elastic modulus
tensor is $\lambda_{VV}= 2\pi B^2/V^3$ corresponding to finite compressibility.
All other components of $\lambda$ vanish, as they should, for a fluid; the two
shear modes are soft.

An orthonormal, smooth, family of single particle states that
span the lowest Landau level is in fact given by theta functions
\cite{Mumford}:
\begin{eqnarray}
\varphi_\ell(x,y)= \frac{(2 \,\tau_2\ B)^{1/4}}{V^{1/2}}\ \sum_{n=-\infty}
^{\infty} e^{i\pi\tau B (\tilde y+n)^2}\,e^{-2i\pi(\phi_2+B/2)(\tilde y+n)}\,
e^{2\pi i(nB+\ell) x}, \label{wave-function}\end{eqnarray} where
$\tilde y =y +\ell/B +\phi_1/B+1/2$  and
$\quad \ell=0,\dots,B-1$.

A little computation shows that the component of the  adiabatic
curvature  in Eq. (6), which corresponds to the $\ell$-th state, is:
\begin{eqnarray} \sum_{i<j} \Omega_{ij}^\ell\  dX_i\wedge dX_j
=\frac{d \tau_1\wedge d\tau_2}{4\tau_2^2}-\frac{2\pi}{B}d \phi_1\wedge
d\phi_2\label{curvature}\end{eqnarray}
(cf. \cite{levay}). The basic  equations of
transport (the main result of this letter) follow from
Eqs.~(\ref{fermions},\ref{q-transport}) and (\ref{curvature}):
\begin{eqnarray}  \langle
\frac{\partial H}{\partial V}\rangle =-2\pi B^2/V^2,\quad \langle
\frac{\partial
H}{\partial\tau_1}\rangle
 = \frac{B}{4\tau_2^2}\  \dot \tau_2,\quad
\langle \frac{\partial H}{\partial\tau_2}\rangle
 = -\frac{B}{4\tau_2^2}\  \dot \tau_1.\label{main-result}
\end{eqnarray}
There is a single transport coefficient related to viscosity and it is
${B}/(4\tau_2^2)$. The interpretation of this result is as follows.
Consider two dimensional Hall fluid on a surface of a cylinder. Compressing
it in the radial direction (or in the axial direction) results in a twist rate
of one boundary circle relative to the other. And vice versa: a shear of the
two
boundary circles results in a compression rate of the radial  direction
and stretching rate of the axial direction.

It may be instructive to compare these transport equations, Eq.
(\ref{main-result}), with the transport equations for the Hall conductance for
Landau levels on the torus. These too follow from
Eqs.~(\ref{basic-transport}) and (\ref{curvature}). Namely,  \begin{eqnarray}
\langle \frac{\partial H}{\partial\phi_1}\rangle
 = 2\pi \dot \phi_2,\quad
\langle \frac{\partial H}{\partial\phi_2}\rangle
 = - 2\pi\dot \phi_1.\label{hall} \end{eqnarray}
The generalized force in this case is the current operator $\frac{\partial
H}{\partial\phi}$, and the generalized velocity $\dot\phi$ is the electromotive
force. By Eq. (\ref{hall}) the conductance of a full Landau level is $2\pi$
in our units  and $e^2/h$ in ordinary units.

Comparing the conductance and
viscosity one sees that the viscosity increases linearly with magnetic field,
while the conductance is independent of it, and that the conductance is
constant on the flux space while the viscosity is $\tau_2$-dependent.

To appreciate the geometric significance of Eq. (\ref{main-result}) we will
look at the space of parameters $(V,\tau,\phi)$ more thoroghly. The modular
group $SL(2,Z)$ acts naturally on this parameter space as a symmetry group
of the family of Landau Hamiltonians $H(V,\tau,\phi)$, Eq. (\ref{hamiltonian}).
To describe this action explicitly let us note first that an element
$M=\left(\begin{array}{cc}a&b\\c&d\\\end{array}\right)\in SL(2,Z)$ provides an
isometry between metrics $g(V,\tau)$ and $g(V,\tau^\prime)$ with
$\tau^\prime=\frac{a\tau+b}{c\tau+d}$ by the formula
$$x\mapsto ax-by,\; y\mapsto -cx+dy.$$
We can therefore consider $M$ as a unitary operator in the Hilbert space. Then
operator $MH(V,\tau,\phi)M^{-1}$ coincides up to a gauge transformation
with $H(V,\tau^{\prime},\phi^{\prime})$, where
$$\phi_1^{\prime}=d\phi_1+b\phi_2,\,\phi_2^{\prime}=c\phi_1+a\phi_2.$$

The last formula shows, in particular, that the family of Landau Hamiltonians
$H(V,\tau,0)$ is $SL(2,Z)$-invariant and the corresponding parameter space is
nothing else but the moduli space of elliptic curves. It is a 2-sphere with
two conical points and one puncture. It can be conveniently represented by
the fundamental domain of $SL(2,Z)$-action on the upper half plane of
complex variable $\tau$, Fig. 2 \cite{Mumford}. It coincides with the
parameter space of flat metrics on a torus of fixed area. This is an
analog of the Aharonov-Bohm flux torus in the theory of Hall conductance.

The geometric significance of the viscosity in Eq. (\ref{main-result}) is now
apparent: the first term in Eq.~(\ref{curvature}) is the invariant area
form on the upper half plane of complex variable $\tau$. Similar to the
Hall conductance, which is  constant in the flux space, the viscosity  is
proportional to the area form in the moduli space of elliptic curves.

In the Hall effect, the conductance is associated with a topological invariant
- Chern number, which is an integer. This integer comes from integrating the
curvature over parameter space. In the case of Eq. (\ref{hall}) this integer is
1. For the viscosity the situation is almost the same. Integration of the
adiabatic curvature over the moduli space (or fundamental domain $F$ of the
group $SL(2,Z)$) gives

\begin{eqnarray}
\frac{1}{2\pi}\int_F {\frac{B}{4}\,\frac{d\tau_1\wedge d\tau_2}{\tau_2^2}} =
\frac{B}{24}, \end{eqnarray}
where we have used the fact that the area of the fundamental domain is $\pi/3$.
Though not an integer in general, this is still a topological invariant -
Chern number (in the orbifold sense) of the ground state bundle on the
parameter space. (It is not an integer because the parameter space is not a
smooth compact manifold in this case).

 \section*{Acknowledgment}
We thank A. Auerbach for discussions. This work is supported in part by the
DFG,
the GIF and by the Fund for the Promotion of Research at the Technion. RS and
PZ acknowledge the hospitality of the ITP at the Technion.

\vfil\eject
\section*{Figure Captions}


\bigskip
Fig.~1. Parallelogram associated with complex parameter
$\tau$.\hfil\break
\bigskip\noindent
Fig.~2. Fundamental domain of $SL(2,Z)$.\hfil\break


\begin{thebibliography}{article}
\bibitem{levay} While writing this paper we learned that the adiabatic
curvature for fixed $\phi_{1,2}=-B/2$
 (the first half of Eq.~(\ref{curvature})) was independently
computed by P.L\'evay who has also calculated the curvature for higher Landau
levels.
P. L\'evay, to appear in J. Math. Phys.\par
\end{thebibliography}
\end{document}